# IAU PLANET DEFINITION: SOME CONFUSION AND THEIR MODIFICATIONS

R. Sarma[a], K. Baruah[b], J. K. Sarma[b*]

a Department of Physics, Hojai College, Hojai-782435, Nagaon, Assam, India
E-mail: sharma.rathin@gmail.com
b Department of physics, Tezpur University, Napam-784028, Tezpur, Assam, India
E-mail: jks@tezu.ernet.in

International Astronomical Union (IAU) has passed the must needed definition of planet in its general assembly held in Prague during August 2006. The definition had to be passed by means of voting. A group of scientists who raised the banner of revolt against the IAU definition has pointed out that the IAU has failed to give an acceptable definition regarding a planet. A brief description of the serious objections found in the definition of planet has been discussed here. In this paper an attempt has also been made to give a new definition of a planet by introducing some modifications to the IAU definition.

## INTRODUCTION

Before the IAU definition[1], a 'planet' was never been scientifically defined. We found the proper concept of a planet for the first time when Copernicus proposed the heliocentric Solar System model[2]. According to him all planets including the Earth move around the Sun. In other words, any celestial body that moves round the Sun is a planet. This type of 'concept' prevailed for a very long time. When Ceres was discovered in 1801 by Giuseppe Piazzi[3] it was categorized as a planet, subsequently it became clear that it was not a planet but a member of the asteroid belt. Same thing happened with the Pluto also[1]. When Tombaugh discovered Pluto in 1930[4], immediately it was placed in the planet list. Pluto enjoyed the status as a planet of the Solar System up to 24th August 2006, the day when IAU gave the new definition of a planet[1]. According to the new definition Pluto loses its planetary status and it becomes a dwarf planet. Of course, on 11th June, 2008 again another resolution has been passed by the IAU where all dwarf planets which are beyond Neptune are termed as 'Plutoids'[5].

## THE CONTROVERSY

The lack of a proper definition of a planet has turned the process of categorizing the planets into a big controversy.

Actually controversy has started from the day when Pluto was discovered. The planetary status of Pluto has been questioned by various scientists for several years[6]. But the topic gained momentum as soon as large Kuiper belt objects like Eris were discovered. After the discovery of Eris, which is slightly larger in size than Pluto[7], the doubt regarding the planetary status of Pluto becomes a hot topic both in scientific community and in media. The process of discovering new solar objects has been going on at a quantum speed and the list of new objects to be categorized has already become too long. If Ceres and Eris are to be counted as planets then it is impossible to restrict the entry of another 42 objects as planets. As a result there will be 53 planets in the Solar System making it more complicated[8]. This inspired the scientists to take up the matter seriously and they started to 'rethink' the definition of a 'planet'.

**THE IAU XXVIth GENERAL ASSEMBLY**

The XXVIth General Assembly of the IAU was held from August 14 to August 25, 2006 in Prague, Czech Republic. The final resolution on the definition of a planet was passed on August 24 by the Assembly, which classified Ceres, Eris and Pluto as dwarf planets, and reduced the number of planets in the Solar System to 8. The voting procedure followed IAU's Statutes and Working Rules. The General Assembly lasted 12 days and had 2412 participants, but only 424 IAU members attended the Closing Ceremony held on 24th August 2006 and took part in the voting process[1].

**THE IAU RESOLUTION**

The IAU resolves that 'planets' and other bodies in our Solar System, except satellites, be defined into three distinct categories in the following way:[1]

(1) Planet: a 'planet' is a celestial body that
   (A) is in orbit around the Sun,
   (B) has sufficient mass for its self-gravity to overcome rigid body forces so that it assumes a hydrostatic equilibrium (nearly round) shape, and
   (C) has cleared the neighborhood around its orbit.

(2) Dwarf planet: a 'dwarf planet' is a celestial body that
   (A) is in orbit around the Sun,
   (B) has sufficient mass for its self-gravity to overcome rigid body forces so that it assumes a hydrostatic equilibrium (nearly round) shape,
   (C) has not cleared the neighborhood around its orbit, and
   (D) is not a satellite.

(3) Small Solar System bodies: all other objects except satellites orbiting the Sun shall be referred to collectively as 'Small Solar-System Bodies'.

**DISCUSSION**

The IAU has given three definitions regarding planets, dwarf planets and small Solar System bodies. But we have some confusions and objections about these definitions.

(a) As a first criterion for planet, it is said that the object should be in the orbit around the Sun. This means that the definition is valid within the Solar System only. We feel that it is not a right move by the IAU. At a time when even common people are aware of the other planetary systems and the 'exoplanets', the definition should be universal one. Among 249 exoplanetary systems, 29 are known to possess more than one planet[9]. Again most of these exoplanetary systems have non-intersecting orbits, with only three exceptions[10]. Therefore, a general validity of the definition of a planet is expected.

As by the word 'star' we never restrict ourselves with the Sun only, similarly by the word 'planet' why should we mean the 8(?) Solar planets only? If the IAU wishes to give a definition of a planet for the Solar System objects only, then it would be better to specify the matter like definition of 'solar planet'. It seems that the IAU was in a hurry to skip the Pluto controversy rather than giving a proper definition of a planet because IAU assembly with 2412 participants ended the matter with voting by only 424 members.

(b) The second criterion indicates the required lower mass limit and shape of the object to be a planet. At hydrostatic equilibrium, an object in the absence of rotation takes the shape of a sphere. The shape or figure of a 'planet' depends upon rotation rate of the body also. Rotation flattens a deformable object somewhat changing its figure to an oblate spheroid[11]. Again to attain a spherical shape, an object does not completely depend on mass but also on the density and compressive strength of the material[10]. Much smaller bodies can also attain round shape via melting or through other form of asteroid differentiation[12]. Additionally, from the round shape of km-sized rubble pile asteroids we know that roundness can also be acquired via violent kinetic events[13]. Other examples of anomalies of shape are that the small icy satellite Mimas (395 km diameter) looks round while the rocky asteroid Vesta (538 km) is clearly non-spherical[14].

The roundness concept is not clear. It is rather confusing. Therefore, the idea of 'nearly round' makes it impossible for any body to judge how much round it should actually be. Again this type of 'roundness' will certainly vary from man to man. Someone may consider 1% deviation from pure roundness as not 'nearly round', while someone else may consider 10% deviation as 'nearly round'. These facts hint that 'roundness' could be connected to the cosmogony of the object rather than to its physical properties only and as such would not be an ideally suited observable. Hence, roundness is not a proper concept at the bottom of the planetary mass scale[15].

(c) The third point is also about lower mass limit. Stern and Lavison (2002) remarked that some bodies in the solar system are dynamically important enough to have cleared out most of the neighboring planetesimals in a Hubble time, while lesser bodies, unable to do so, occupy transient unstable orbits, or are prevented in mean motion resonances or satellite orbits[10]. Though this is an important criterion to categorize the celestial objects in different groups, we feel that the words that are used to describe this by IAU, are not appropriate which may lead to some confusion. According to the statement the object should 'clear its neighborhood'. But how could one choose which are the objects that should be considered as neighbor for a particular object? Someone may consider Mercury as a neighbor of the Earth.

**SOME OTHER DEFINITIONS**

Before the inception of the IAU definition, various scientists throw light on this from various angles. The root cause behind these definitions revolves round the Pluto controversy.

Gibor Basri of Astronomy Department in University of California gives a definition of a planet which is purely mass-based. He describes a planet as - 'Planet: a planemo that orbits a fusor. Planemo: a round non- fusor. Fusor: an object capable of core fusion.'[14]

Michel E. Brown of California Institute of Technology, though prefers to set lower mass limit ('clearing neighborhood' idea) to define a planet, has added culture with it. According to him (a) a planet is an object that is massive enough to clear planetesimals from its orbital neighborhood, and which is a part of the empirically defined distinct group of low mass stellar companions with masses lower than about 5 Jupiter masses and (b) in the Solar System a planet is any of the nine historical planets plus any newly found objects bigger than the smallest of these[14].

Steven Soter of Department of Astrophysics in American Museum of Natural History proposes a definition which is a combination of both cosmogony and mass. He defines it as follows. (i) A 'primary' body which is a star or substar formed by core accretion from an interstellar cloud, not by secondary accretion from a disk. (ii) A 'substar' is a body which is less than 80 Jupiter masses, the lower limit for stellar hydrogen fusion. (iii) A 'planet' is an end product of secondary accretion from a disk around a primary body. (iv) An 'end product' of disk accretion is a body containing more than 100 times the mass of all other bodies that share its orbital zone. (v) Two bodies share an 'orbital zone' if their orbits cross a common radial distance from the primary and their products are non-resonant and differ by less than an order of magnitude.[10]

Bojan Pecnik of Croatian National science Foundation and Christopher Broeg of University of Bern define a planet with a concept for a global static critical core mass[16]. According to them a planet will have a core which is supercritical within the appropriate manifold. They have not called an object a planet if it is not capable to retain its envelope (volatiles) when connected to vacuum.[15]

Alan Stern and Harold F. Levison of Department of Space Studies in Southwest Research Institute favor a definition which is based on upper and lower mass limit. They defined a planetary body as any body in space that satisfies the following testable upper and lower bound criteria on its mass. If isolated from external perturbations (e.g. dynamical, thermal), the body (1) be low enough in mass that at no time (past or present) can it generate energy in its interior due to any self sustaining nuclear fusion chain reaction (else it would be a brown dwarf or a star) and also (2) be large enough that its shape becomes determined primarily by gravity rather than mechanical strength or other factors (e.g. surface tension, rotation rate) in less than a Hubble time, then the body would on this timescale or shorter reach a state of hydrostatic equilibrium in its interior. According to Stern and Levison a planet is any planetary body on a bound orbit around a single or multiple star system[17]. We have not discussed these definitions individually because the main objective of this paper is to discuss the IAU definition only.

**SOME MODIFICATIONS**

(1) We would like to replace the word 'the Sun' with 'a star' or 'a substar'[10] from the first point of the planet definition given by the IAU. Like the Solar System, almost all exoplanetary system consists of a central star. For exoplanetary system, the presence of a central star is revealed by the methods by which the exoplanets are being detected. As distant planets are extremely faint, most methods of detecting exoplanets are indirect, in the sense that the planet is detected through its influence on the star that it orbits[11]. Some exoplanets are found without having central stars which are commonly known as 'free-floating planets'[18]. Two possibilities for these types of body formation are given. One of them is that they formed in planetary systems around stars and were subsequently ejected from the system by interaction with other massive bodies in the system. The other probability is that these objects are formed in isolation, or at least were not originally bound to a star[19]. In either case, the entry of these types of objects into planet category will be restricted by the criterion of having a central star.

(2) As we have discussed above, the second criterion is problematic and confusing. By 'shape' (nearly round) it is very hard to draw a dividing line between a planet and a non-planet. As the third criterion (clear neighborhood) also deals with mass and is more acceptable, we propose that this second criterion should completely be removed from the planet definition.

(3) The third criterion is suitable one as it reveals important aspects of the process that formed the Solar System. Planet formation by the process of accretion of small bodies, known as planetesimals, is widely accepted. The accretion process led to the formation of embryo planets that as they grew in size and acquired more powerful gravitational fields, went to a process of runway accretion in which the size of a few of them detached from the rest of the bodies of their neighboring zones. These protoplanets were able to clean the population that had close encounters with them. The remaining planetesimals were finally incorporated to the planets or scattered to other regions[20].

The measure of extent to which a body dominates the other masses in its orbital zone is a physically significant parameter[10]. Over many orbital cycles, a large body will tend to cause small bodies either to accrete with it, or to be disturbed to another orbit. For dominating (clearing) the neighbors, the condition require is that the surface escape velocity (the maximum velocity that one body can impart on another through gravitational interaction) should be greater than the local escape velocity from the central star. A body of such mass will be able to scatter other small bodies beyond the gravitational influence of the star. This criterion is met when the ratio of the planet to central star mass is greater than the ratio of the planet's distance from the star to the radius of the star[14]. To measure the extent to which a body scatters smaller masses out of its orbital zone in a Hubble time, Starn and Levison derived a parameter $\Lambda$, where $\Lambda = kM^2/P$, k being approximately a constant, and M and P are mass and orbital period of the scattering body respectively. Starn and Levison have shown that the values of $\Lambda$ for the planets are remarkably higher than those of the non planets[17].

To represent this type of mass dominance by the massive objects in their orbits the term 'clear neighborhood' is not the best fit, because 'clear neighborhood' will mean a totally empty (clean) unspecified region around the orbits. But practically the situation is not like that. There may be some small mass objects but the 'planet' mass must dominate them in a definite region. This definite region is the 'orbital zone' of the object. Therefore, we feel that replacement of 'clear neighborhood' by 'clear most of the mass from the orbital zone' will make the statement more meaningful to all. So we can modify the definition of a 'planet' as follows:

A 'planet' is a celestial body that
(a) is in orbit around a central star and
(c) has cleared most of the mass from its orbital zone.

**DWARF PLANETS AND SMALL SOLAR SYSTEM BODIES**

IAU has introduced two new groups of objects, one is 'Dwarf Planets' and the other is 'Small Solar System bodies'. Again on 11th June, 2008 another resolution has been passed by the IAU where all dwarf planets which are beyond Neptune are termed as 'Plutoids'.

It is not yet clear whether dwarf planet status is, like planet status, a sole defining category, or whether dwarf planets also retain their previous minor body classifications such as 'asteroids'. The only criterion remains (other than orbiting the Sun) for dwarf planets to be different from the small bodies, as given by IAU is the shape i.e. 'roundness'. In this case also it will be difficult to draw a dividing line between dwarf planets and small solar system bodies. As we have discussed above, the presence of small objects with round shape will definitely create confusion.

Again according to IAU, small Solar System bodies will include all objects orbiting the Sun except the satellites. Does it mean the inclusion of all planets and dwarf planets in this group? From the given definition this is not clear at all. We, therefore, strongly feel that creation of these groups (dwarf planets, small solar system bodies and plutoids) is confusing and unnecessary. The object that fails to meet the required criterion to be a planet should remain as an object of the region from where it originates. We have the experience of Ceres. Ceres, the largest asteroid and discovered first (on January 1, 1801) was originally classified as a planet, and kept this status until we discovered that it was just the largest of a class of objects we now call 'asteroids'. Similarly, if Pluto is not a planet, it obviously is a 'Kuiper belt object'.

**CONCLUSION**

The definition of a planet is very important for scientific advancement. It has much more importance for common people and for school level students as well. So the definition should carry an easily understandable clear (not confusing) scientific meaning. Our proposed definition will make an attempt to meet this objective. According to the proposed definition there will be eight planets in our Solar System, Pluto will be excluded from the planet list. Eris, Pluto and MakeMake[21] should be regarded as the members of the Kuiper belt object.